# Experimental Comparison of PAM-8 Probabilistic Shaping with Different Gaussian Orders at 200 Gb/s Net Rate in IM/DD System with O-Band TOSA

*Md Sabbir-Bin Hossain[(1,2)]\*, Georg Böcherer [(1)], Youxi Lin[(1)], Shuangxu Li [(1)], Stefano Calabrò[(1)], Andrei Nedelcu[(1)], Talha Rahman[(1)], Tom Wettlin[(2)], Jinlong Wei[(1)], Nebojša Stojanović[(1)], Changsong Xie[(1)], Maxim Kuschnerov [(1)], and Stephan Pachnicke[(2)]*

[(1)] Huawei Technologies Duesseldorf GmbH, Munich Research Center, 80992, Munich, Germany
[(2)] Chair of Communications, Christian-Albrechts-Universität zu Kiel, 24143 Kiel, Germany
**\***E-mail: sabbir.hossain@huawei.com

**Abstract** *For 200Gb/s net rates, cap probabilistic shaped PAM-8 with different Gaussian orders are experimentally compared against uniform PAM-8. In back-to-back and 5km measurements, cap-shaped 85-GBd PAM-8 with Gaussian order of 5 outperforms 71-GBd uniform PAM-8 by up to 2.90dB and 3.80dB in receiver sensitivity, respectively.*

**Introduction**

In the coming years, data center traffic is expected to grow exponentially due to the cloud services such as video-on-demand, cloud computing, gaming, and internet-of-things (IOT). To date for short reach (below 10 km), intensity modulation with direct detection (IM/DD) is preferred over coherent systems, due to the advantages like low power consumption, simpler transceiver design as well as smaller footprint [1]. For operating close to the Shannon limit and achieving finer granularity probabilistic shaping (PS) is vastly explored and even implemented in commercial coherent system-based products for long-haul transmission. Due to the availability of high bandwidth opto-electronic components [2,3] PS is also being explored for IM/DD based short reach applications [4-7]. In our recent publications, with cap-shaped PS with Gaussian order of 2, in term of receiver sensitivity, we show a performance improvement of up to 1 dB [4] and 3.5 dB [5] in C and O-band, respectively. Furthermore, with a combination of PS and geometric shaping (GS) up to 1.1 dB performance gain has been reported [6].

This work is a follow-up on our previous paper [5], where we have demonstrated that 2nd order Gaussian shaped Maxwell Boltzmann (MB) distributed cap-shaped PS outperforms cup-shaped PS in the considered scenarios. In this work, we evaluate MB distributions with different Gaussian orders [8] (Fig. 1) for cap-shaped PS only. The implementation technique is described in the next section. We compare experimentally this scheme with uniform 8-level pulse amplitude modulation (PAM). The experimental setup is based on a transmitter optical subassembly (TOSA), which is next generation feasible component on the transmitter side. The experimental results indicate that for PS PAM-8, higher-order Gaussian MB distributions outperform signalling without shaping or lower-order MB distributions. Furthermore, with 7% hard decision (HD) forward error correction (FEC) with bit error rate (BER) threshold of $3.8\times10^{-3}$, 85 GBd cap-shaped PAM-8 with an entropy (H) of 2.5492 bit/symbol improves the receiver sensitivity by up to 3.80 dB for the case of 5 km transmission compared to 71 GBd uniform PAM-8.

**Cap PS With Different Gaussian Orders**

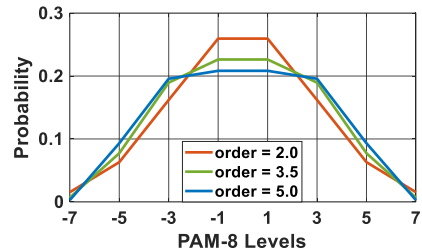

Fig. 1 Gaussian order variant

The spectral efficiency of a transmission system with probabilistic amplitude shaping (PAS) [9] is given by

$$\text{SE} = \text{H}(P_X) - m(1 - R_{\text{FEC}}) \quad (1)$$

where $P_X$ is the input distribution $\text{H}(P_X)$ the entropy in bits, $m$ the number of bit-levels, i.e., $m = 3$ for PAM-8, and where $R_{\text{FEC}}$ is the FEC rate. The PS overhead in bits is $m - \text{H}(P_X)$. For fixed net bit rate and FEC overhead, the PS overhead is a function of the baudrate. Most commonly, the MB distribution

$$P^{\text{MB}}(x) \propto \exp(-\nu x^2) \quad (2)$$

Is chosen, as it minimizes average power. The parameter $\nu$ is chosen according to the desired overhead. Consequently, for each PS overhead,

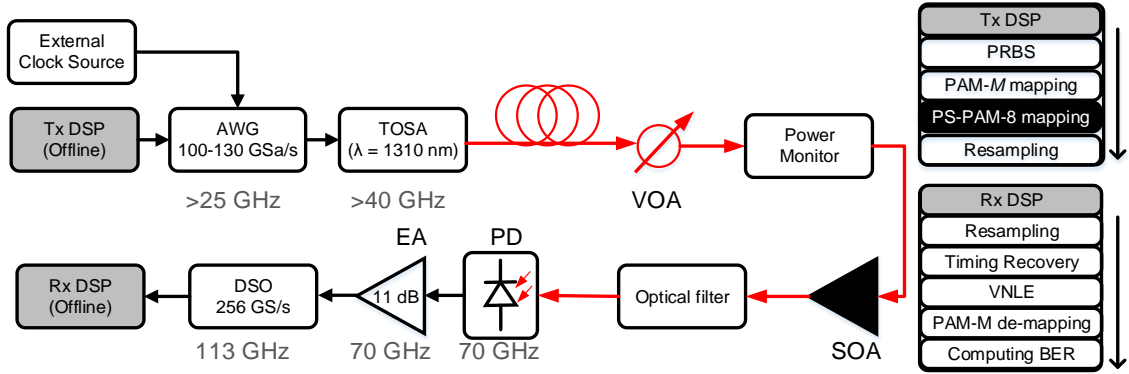

AWG: arbitrary waveform generator; TOSA: transmitter optical sub assembly; VOA: variable optical attenuator;
SOA: semiconductor optical amplifier; PD: photodetector; EA: electrical amplifier; DSO: digital storage oscilloscope
**Fig. 2:** Schematic of the IM/DD transmission system with transmitter and receiver-side offline DSP.

there is only one MB distribution. For having a whole family of distributions, we therefore introduce as additional parameter the Gaussian order $\alpha$. The family of distributions then becomes

$$P^{\nu,\alpha}(x) \propto exp(-\nu|x|^{\alpha}) \qquad (3)$$

For each overhead, we now have several candidate distributions depending on the order $\alpha$ and we can experimentally determine which distribution performs best.

**Experimental Setup and DSP**
The experimental setup is represented in Fig. 2. Below each electro-optical component, its 3-dB bandwidth (BW) is indicated. The offline transceiver digital signal processing (DSP) is also depicted as a sequence of steps. A pseudorandom binary sequence (PRBS) was generated and mapped to a PAM-8 levels' array. Similarly for PS PAM-8, random symbols were drawn from a cap MB distribution with desired entropy. Considering 7% HD-FEC, targeting 200 Gbit/s net bit rate, a symbol rate of 71 GBd was used for uniform PAM-8, while PS PAM-8 with symbol rates of 80, 85, and 90 GBd was generated with entropies of 2.6963, 2.5492, and 2.3867 bit/symbol, respectively. The generated symbol sequence was then pulse-shaped with a root-raised cosine filter with a roll-off factor of ≈ 0.4.

The quantized data was then fed to an arbitrary waveform generator (AWG) with a 3-dB BW of >25 GHz. The AWG could be operated between 100 and 130 GSa/s with the help of an external clock source. The sampling frequency was varied based on the transmitted symbol rate to accommodate the interpolated signal with a fixed 0.4 roll-off factor. For 71, 80, 85, and 90 GBd the sampling frequencies were set to 100, 107, 113, and 120 GSa/s, respectively. Afterwards the analog signal was fed to a TOSA, where it is amplified with an integrated driver amplifier and modulated with an electro-absorption modulated laser (EML).

The experimental measurements were conducted both for the B2B case and for a transmission distance of 5 km over standard single-mode fiber (SSMF). A variable optical attenuator (VOA) was used after the fiber link to regulate the received optical power (ROP). A commercial receiver optical sub assembly (ROSA) would contain a PIN photodetector (PD) and a phase-matched transimpedance amplifier (TIA). However, due to the unavailability of a ROSA during the experiment, the receiver architecture was modified with a 70 GHz PIN PD accompanied by a 70 GHz 11 dB fixed gain electrical amplifier (EA). Considering the experimental setup and the operation point of the EML, the best B2B performance was achieved at an optical transmitter power of ≈ -3.8 dBm, which is relatively low to investigate the performance of the link without a ROSA. To compensate for that low power at the PD, the received optical signal was pre-amplified by a semiconductor optical amplifier (SOA) and passed through an optical filter (width 2 nm) to suppress the amplified spontaneous emission (ASE) noise.

Afterwards, the received electrical signal was quantized and captured with a digital storage oscilloscope (DSO), which operates at 256 GSa/s with a 3-dB BW of 113 GHz. Data was captured for different values of ROP measured at the input of the SOA and adjusted by the VOA. Offline processing at the receiver is based on 0.5 million samples. For estimating and compensating sampling frequency offset and jitter, timing recovery is performed [10]. After time synchronization, the signal is resampled to 1 sample per symbol and processed with a Volterra nonlinear equalizer (VNLE). A memory length of 311 taps was used for linear equalization to minimize the effect of reflections from the discrete component setup, copper cables, and connectors. In addition to linear taps, 2[nd] and 3[rd] order kernels of VNLE with a memory length of 11 were used to compensate for nonlinearities introduced by AWG, EML, SOA, EA, and square-law detection. Afterward, symbol decision was performed on the PAM-8 symbols. The symbol decision thresholds were optimized based on the distribution probabilities of each level. Afterward, the symbols were de-mapped to bits and the BER was calculated.

**Transmission Results and Discussions**

shaped PAM-8 80 GBd with Gaussian order of 2, since the overall channel BW becomes critical and a comparatively high number of symbols on the outer levels makes the overall performance

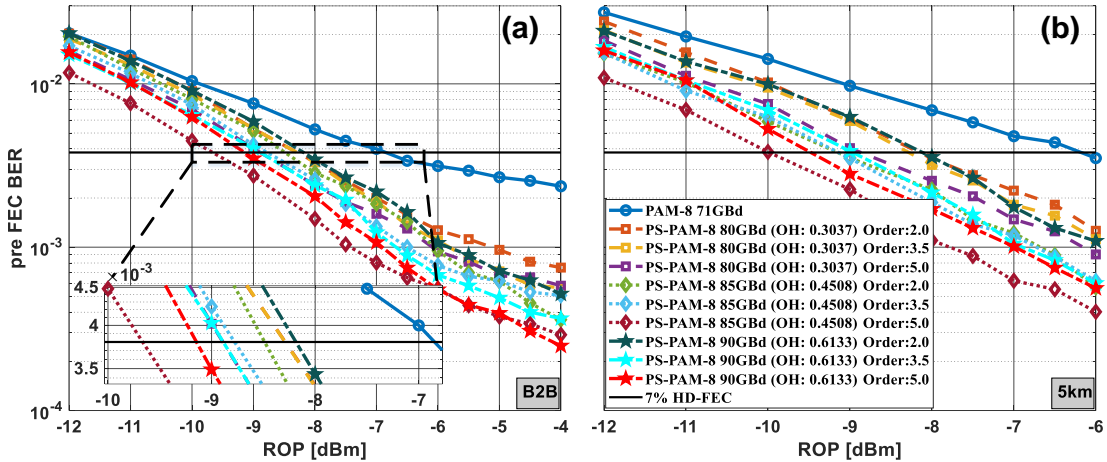

**Fig. 3:** Experimental results of uniform and probabilistic shaped PAM with different Gaussian order for B2B (a) and 5 km(b).

The experimental results are presented in Fig. 3 in terms of pre FEC BER variation over ROP. Results measured for B2B and 5 km transmission and are shown in Fig. 3(a) and (b), respectively. As a performance threshold criterion, 7% HD-FEC is considered. A solid line with circle markers represents uniform 71 GBd uniform PAM-8. The dashed lines with square markers represent cap-shaped 80 GBd PAM-8, and dotted lines with diamond markers indicate cap-shaped 85 GBd PAM-8 with Gaussian order of 2, 3.5 and 5, respectively. Similarly, dashed-dotted lines with pentagon markers represent cap-shaped 90 GBd PAM-8.

Evaluating the results in terms of receiver power sensitivity, uniform PAM-8 has reached the HD-FEC threshold at -6.84 dBm. On the other hand, cap-shaped PAM-8 80 GBd with Gaussian order 2 and 3.5 reaches the HD-FEC threshold at -8.3 dBm, while with Gaussian order 5, the performance is improved by up to 0.6 dB. The performance with cap-shaped 85-GBd PAM-8, with Gaussian order 3.5 performs almost similar to the cap-shaped 80 GBd PAM-8 with Gaussian order 5. Furthermore, cap-shaped 85 GBd PAM-8 with 2.5492 bit/symbol with Gaussian order of 5 performs 0.8 dB better compared to the cap-shaped 80 GBd PAM-8 with 2.6963 bit/symbol with Gaussian order 5, and approximately 2.9 dB better compared to the 71 GBd uniform PAM-8. The comparison is particularly interesting for cap-shaped PAM-8 90 GBd with 2.3867 bit/symbol for different Gaussian orders. For instance, for the Gaussian order of 2, the performance is very similar (only slightly worse) compared to cap degradation. As expected, the performance and is improved by up to 0.8 and 1 dB with Gaussian order 3.5 and 5, respectively, as the outer symbols become less probable.

For 5 km transmission case, uniform PAM-8 shows ≈ 0.6 dB penalty compared to B2B case, while transmission with PS PAM-8 shows negative penalty (about -0.3 dB) with Gaussian order 5. This also indicates that accumulated dispersion is much more severe and contributes to the inter-symbol interference (ISI) on uniform symbol distribution compared to the fewer symbols on the outer levels.

**Conclusion**

We have shown an experimental comparison between uniform PAM-8 with cap-shaped Maxwell-Boltzmann distributed PAM-8 with different symbol rates and Gaussian order of 2, 3.5, and 5 for each symbol rate. A net bit rate of 200 Gb/s was targeted and adequate overhead for a 7% HD-FEC was allocated. While comparing uniform PAM-8 and PS PAM-8, PS PAM-8 85 GBd with Gaussian order of 5 performs the best and outperforms uniform PAM-8 by around 2.9 dB. For 5 km transmission, the power sensitivity performance is improved up to 0.9 dB over B2B scenario.